\documentclass[runningheads]{llncs}

\usepackage{eccv}

\usepackage{eccvabbrv}

\usepackage{graphicx}
\usepackage{booktabs}
\usepackage{multirow}
\usepackage{xcolor}
\usepackage[accsupp]{axessibility}  

\usepackage{hyperref}

\usepackage{orcidlink}

\begin{document}

\title{A Novel Approach to Linking Histology Images with DNA Methylation} 


\author{Manahil Raza\inst{1} \and
Muhammad Dawood\inst{1} \and
Talha Qaiser \inst{1} \and
Nasir M. Rajpoot \inst{1,2}}

\authorrunning{M.Raza et al.}

\institute{The Tissue Image Analytics Centre, Department of Computer Science, University of Warwick, Coventry, UK \and
Histofy Ltd, Coventry, UK 
\email{\{manahil.raza,muhammad.dawood,talha.qaiser,n.m.rajpoot\}@warwick.ac.uk}}

\maketitle

\begin{abstract}
  DNA methylation is an epigenetic mechanism that regulates gene expression by adding methyl groups to DNA. Abnormal methylation patterns can disrupt gene expression and have been linked to cancer development. To quantify DNA methylation, specialized assays are typically used. However, these assays are often costly and have lengthy processing times, which limits their widespread availability in routine clinical practice. In contrast, whole slide images (WSIs) for the majority of cancer patients can be more readily available. As such, given the ready availability of WSIs, there is a compelling need to explore the potential relationship between WSIs and DNA methylation patterns. To address this, we propose an end-to-end graph neural network based weakly supervised learning framework to predict the methylation state of gene groups exhibiting coherent patterns across samples. Using data from three cohorts from The Cancer Genome Atlas (TCGA) - TCGA-LGG (Brain Lower Grade Glioma), TCGA-GBM (Glioblastoma Multiforme) ($n$=729) and TCGA-KIRC (Kidney Renal Clear Cell Carcinoma) ($n$=511) - we demonstrate that the proposed approach achieves significantly higher AUROC scores than the state-of-the-art (SOTA) methods, by more than $20\%$. We conduct gene set enrichment analyses on the gene groups and show that majority of the gene groups are significantly enriched in important hallmarks and pathways. We also generate spatially enriched heatmaps to further investigate links between histological patterns and DNA methylation states. To the best of our knowledge, this is the first study that explores association of spatially resolved histological patterns with gene group methylation states across multiple cancer types using weakly supervised deep learning. 
  
  \keywords{Computational Pathology \and DNA Methylation \and Graph Neural Networks}
\end{abstract}

\section{Introduction}
\label{sec:intro}
DNA methylation is a crucial epigenetic mechanism in cancer biology that regulates gene expression in cellular processes \cite{moore2013dna}. Aberrant DNA methylation can lead to the activation or silencing of oncogenes critical for cell growth and is one of the most common molecular changes observed in cancer cells \cite{esteller2005aberrant,kulis2010dna}. Hypo-methylation and hyper-methylation refer to DNA methylation levels that are, respectively, lower or higher than those found in standard DNA \cite{ehrlich2002dna}. These epigenetic changes can serve as valuable biomarkers for early cancer detection, prognosis, and targeted therapies \cite{kaminska2019prognostic}. High-throughput DNA methylation assays can produce detailed, genome-wide and high-resolution DNA methylation profiles \cite{zheng2020whole}. The Cancer Genome Atlas (TCGA) project has utilized this technology to profile DNA methylation across more than 10,000 cancer samples \cite{heyn2012dna}. However, despite these advancements, such data is not widely available. Moreover, the turnaround time for methylation-based diagnostic testing can take up to several weeks resulting in significant delays in diagnosis and treatment \cite{hoang2024prediction}. 

Conversely, when cancer is suspected in a patient, routinely stained histology slides are widely available and their analysis continues to be the \emph{gold standard} in clinical cancer diagnostics \cite{khened2021generalized}. Histology slides and their digitized counterparts, whole slide images (WSIs), not only allow pathologists to examine cellular and subcellular structures but also enable the development of deep learning algorithms to identify and mine patterns in these images. However, the inherent characteristics of WSIs, including their varying magnification levels, high-resolution and gigapixel size (an uncompressed WSI can consume up to 100 gigabytes of memory \cite{holzinger2017towards}) present considerable challenges for automated processing \cite{hosseini2023computational}. Consequently, most approaches resort to patch-based analysis, where a WSI is split into hundreds or even thousands of patches and each patch is processed independently by a deep convolutional neural network (CNN). This method poses additional challenges: splitting WSIs into patches results in a loss of important contextual information. Additionally, the ground truth labels are generally available at the WSI level and are typically assigned to all the extracted patches since obtaining patch-based labels is expensive, laborious, and time-consuming \cite{tang2023multiple}. 

Graph Neural Networks (GNNs) are able to overcome the aforementioned challenges associated with patch-based methods by capturing the spatial relationships between tissue structures and cells and allowing the hierarchical modeling of histopathological images \cite{brussee2024graph}. Moreover, GNNs are able to model the entire WSI as a graph, which alleviates the need of obtaining patch-level labels. Deep learning (DL) models including GNNs have significantly enhanced Computational Pathology (CPath) workflows by automating WSI analysis and discovering features that are not visibly apparent \cite{meng2023clinical, hosseini2024computational,jahanifar2023domain}. These algorithms have made notable progress in tasks such as cancer grading \cite{raza2023mimicking, lu2022slidegraph+}, tumor segmentation \cite{bashir2024consistency}, the prediction of molecular pathways, \cite{bilal2021novel}, mutations \cite{jang2020prediction}, survival analysis \cite{graham2024conic} and cancer prognosis \cite{chen2020pathomic}. 

We posit that manipulating DNA methylation patterns with histopathology images can enhance early diagnosis, optimize treatment options, and improve epigenetic mechanistic understanding \cite{yu2024cancer}. Gevaert \etal \cite{gevaert2015pancancer} employed MethylMix \cite{gevaert2015methylmix,cedoz2018methylmix} to identify consistently hypo-methylated or hyper-methylated genes across multiple cancer types. A few methods have been proposed to investigate the relationship between DNA Methylation and histopathology images. Hong \etal \cite{hoang2024prediction} predicted DNA methylation beta values and used the predictions as part of a workflow to classify tumor types, however their analysis was only restricted to central nervous system (CNS) tumors. Zheng \etal \cite{zheng2020whole} classified genes into high and low methylation states using histopathology images. However, they employed hand-crafted features coupled with classical machine learning methods and did not identify spatially resolved histological patterns associated with methylation. To the best of our knowledge, no previous study has utilized deep learning to link histology and DNA methylation states for multiple tumor types. In this study, we propose a graph neural network based workflow, SlideGraph$^{methyl}$, to predict gene group-level differential methylation (DM) states to better understand the associations between tissue WSIs and underlying DNA methylation states. To the best of our knowledge, ours is the first method to use deep learning for the prediction of group-level methylation states across multiple cancer types. 

\begin{figure}[th]
\centering
\includegraphics[width=\textwidth]{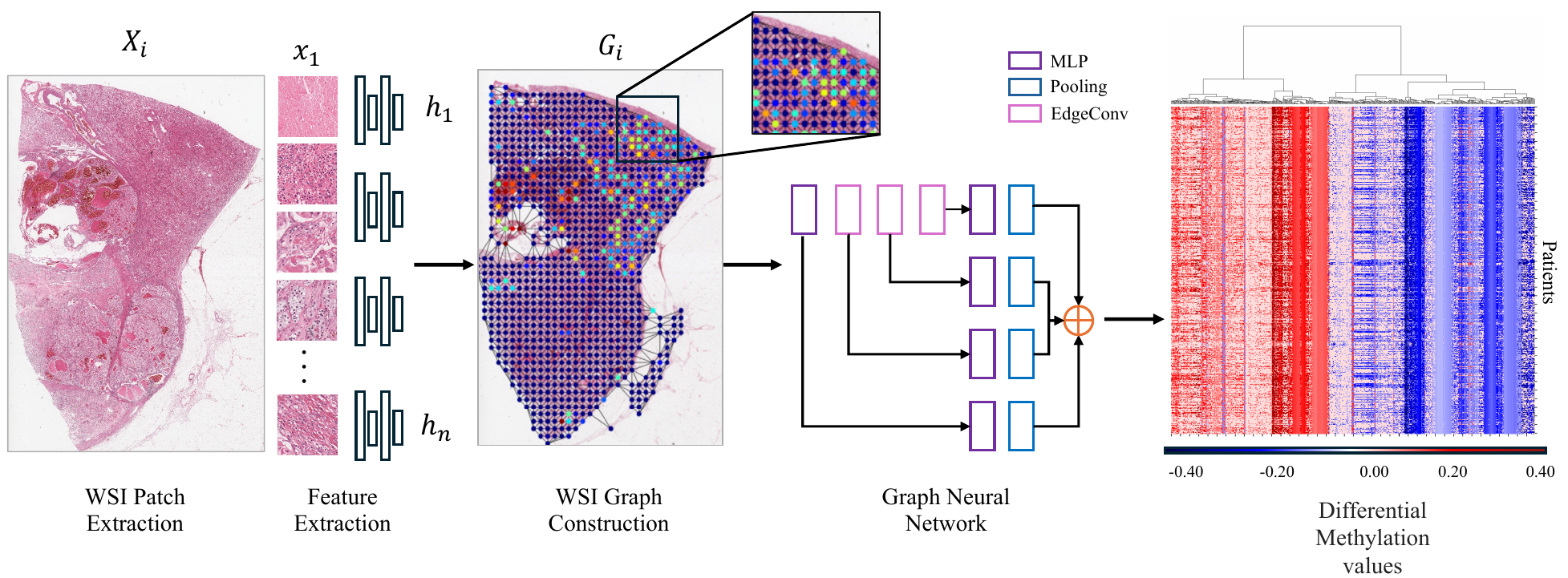}
\caption{The proposed pipeline of SlideGraph$^{methyl}$ for the prediction of the gene group-level methylation status from WSIs. We extract feature representations for the WSI patches to construct a WSI-level graph. This is then fed into a graph neural network to predict the methylation state for a gene group. The hierarchically-clustered heatmaps illustrate the differential methylation (DM) values obtained using MethylMix which serve as the ground truth for this classification problem.}
\label{workflow}
\end{figure}

\section{Methods}
\subsection{Data} 
In this study, we utilized data from three cohorts provided by TCGA \cite{TCGA}, specifically TCGA-LGG (Brain Lower Grade Glioma), TCGA-GBM (Glioblastoma Multiforme) and TCGA-KIRC (Kidney Renal Clear Cell Carcinoma). We limited the analyses to these selected cohorts as a proof-of-concept and preliminary study.  The TCGA-LGG and TCGA-GBM datasets were combined to create the TCGA-GBMLGG dataset, which was used for our study of gliomas. Hematoxylin and Eosin (H\&E) stained WSIs of all patients used in the study can be downloaded from the National Cancer Institute (NCI) Genomic Data Commons (GDC). The DNA methylation data was obtained from the Infinium Human Methylation 450K DNA methylation data from NCI GDC. For the TCGA-GBMLGG dataset, features were extracted from 1,320 histology WSIs belonging to 729 patients with available tissue slides and methylation data. Similarly, for the TCGA-KIRC dataset, features were extracted from 518 WSIs of 511 patients.

\subsection{Pre-processing and Ground Truth for DNA Methylation Data}
\begin{figure}[t]
\centering
\includegraphics[width=\textwidth]{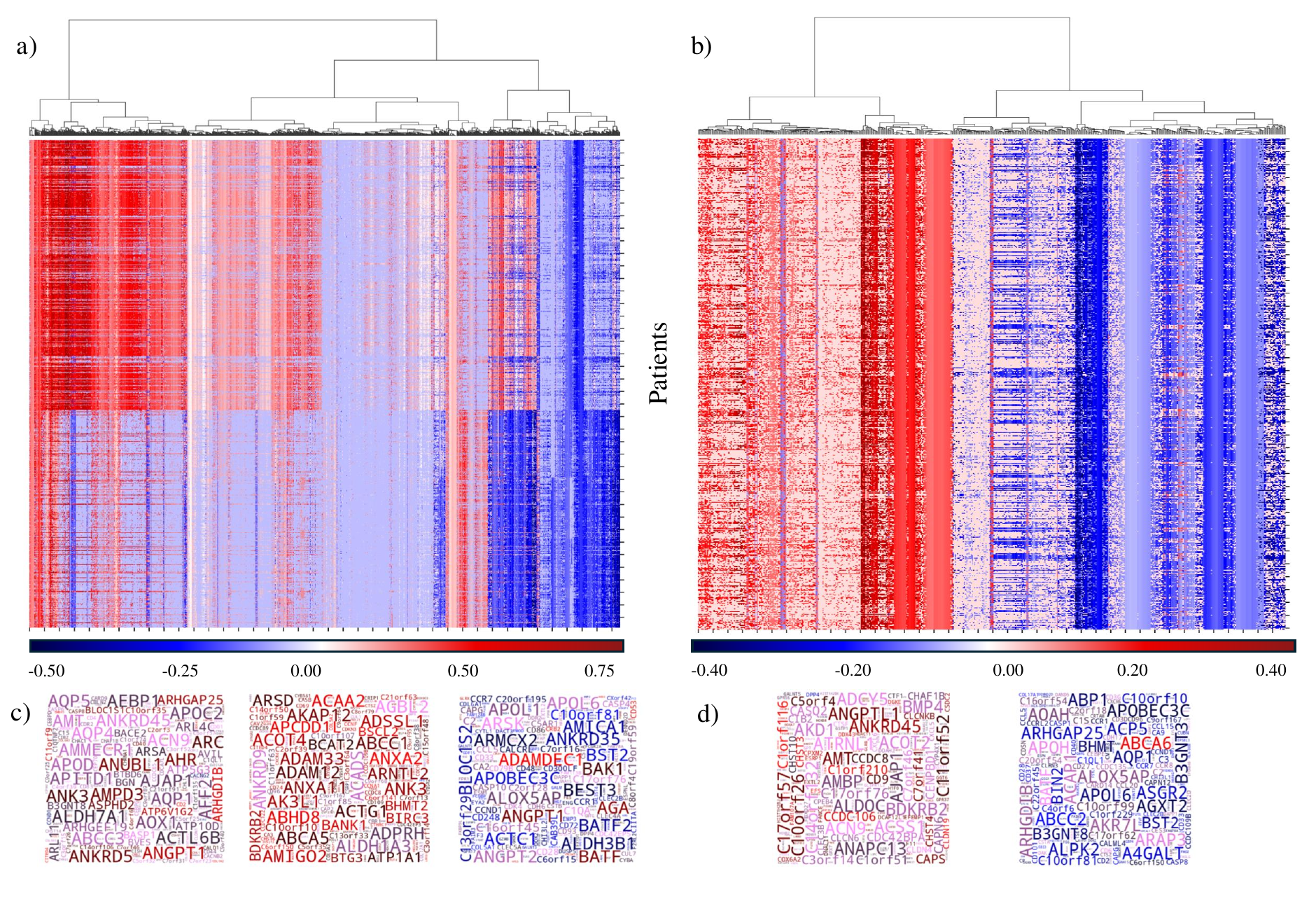}
\caption{Results of hierarchical clustering of the differential methylation (DM) values values for TCGA-GBMLGG (a) and TCGA-KIRC (b). The dendrograms illustrate the clustering of genes (x-axis) based on the DM values for patient cohorts (y-axis). The word-clouds represent the genes in each gene group where red, blue and pink colors indicate hyper-methylated, hypo-methylated and normally-methylated genes, based on their median values across the patient cohorts TCGA-GBMLGG (c) and TCGA-KIRC (d). The font sizes are not representative of anything in particular.} 
\label{data}
\end{figure}

In this study, we predicted differential methylation states for gene groups across a patient cohort. To generate patient-level ground truth, we utilized the MethylMix package  to identify cancer-associated DNA methylation driver genes that are predictive of transcription and exhibit differential methylation in comparison to normal samples \cite{gevaert2015methylmix}. MethylMix calculates differential methylation values which quantify the difference between normal and abnormal methylation states. \cite{cedoz2018methylmix}. The results were organized into a matrix that classifies genes into three categories based on these DM-values: hyper-methylated (positive values), hypo-methylated (negative values), and normally-methylated (zero values) across a patient cohort. Figs. \ref{data} a,b. illustrates the results of hierarchical clustering of the DM values for the TCGA-GBMLGG (1147 genes) and TCGA-KIRC cohorts (519 genes). 

We decided to employ hierarchical clustering to group the genes based on the similarity of their methylation patterns. The dendrograms shown in Figs.\ref{data} a,b represent the results of this clustering process, where the vertical height of the branches in the dendrograms corresponds to the degree of dissimilarity between the gene groups. For the TCGA-GBMLGG cohort, the dendrogram revealed three distinct gene groups, while two groups were identified for the TCGA-KIRC cohort. The genes present in each group for both cohorts are illustrated in Figs.\ref{data} c,d. It is unsurprising that a single gene group seems to contain a majority of either hyper-, hypo- or normally methylated genes. After assigning genes to the groups, we computed the average DM-value for each gene group. Subsequently, in line with Zheng \etal  \cite{zheng2020whole}, we employed a Gaussian mixture model to binarize these DM-values, thereby converting this into a binary classification problem. The ground truth for this classification, therefore, was derived using the MethylMix package, consisting of these binarized gene group-level DM-values as illustrated in Fig.\ref{workflow}. 


\subsection{WSI Analysis Pipeline}
In this study, we propose a computational workflow that we call SlideGraph$^{methyl}$ based on GNNs to predict the group-level methylation states for each patient as illustrated in Fig.\ref{workflow}. The tissue region of each WSI, denoted as $X_i$, was split into non-overlapping patches $X_i = \{x_1, x_2 \ldots x_N\}$  of size $1024 \times 1024$ pixels at a resolution of $0.50$ microns-per-pixel (MPP). 

The CTransPath encoder \cite{wang2022transformer}, pretrained on unlabeled histopathology images, was used to extract feature representations $h_i \in \mathbb{R}^{768}$ for each image patch $x_i$ at location $m_i$. We utilized a hybrid CNN and transformer encoder backbone which combines local feature extraction capability from CNNs and global attention from transformers. We then constructed graph representation, $G = (V, E)$ where $V$ and $E$ denote sets of vertices and edges for each WSI using the feature representations, where each node in the graph $v_i$ corresponds to a patch and can be represented as $v_i = (h_i, m_i)$.  Edges were calculated by connecting neighboring nodes (patches) through Delaunay triangulation, with connections only formed between nodes that were less than 4000 pixels apart (found through empirical evaluation).

We employed a Graph Neural Network (GNN) to process the graph representation of a WSI and provide node level predictions and WSI-level prediction scores. The GNN employs multiple EdgeConv layers to iteratively update a node's feature representations by aggregating information from neighboring nodes. The updated feature representation for node $v_i$ at a layer $l$ with neighbors $\mathcal{N}(i)$ can be described as:
\begin{equation}
h_i^{(l)} = \sum_{j \in \mathcal{N}(i)} \phi^{(l)}(h_i^{(l-1)}, h_j^{(l-1)} - h_i^{(l-1)}; \theta_{l})
\end{equation}
where $\phi$ is a multilayer perceptron (MLP) with learnable weights $\theta_{l}$ and $l= 1,...,L$ with $L$ being the total number of layers in the GNN. 

The updated node representations are then passed through corresponding MLP to produce node-level predictions at layer $l$ as $f_{l}(v_i)=f(h_i^{(l)})$. These node-level predictions are aggregated across all layers, to generate patch-level predictions :
\begin{equation}
    f(v_i)= \sum_{l \in {L}}f_l(v_{i})
,\end{equation} which are further aggregated to generate WSI-level prediction scores, providing an overall methylation state for the patient as follows:

\begin{equation}
    F(G;\theta)=\sum_{i \in {V}}f(v_i).
\end{equation}
For patients with multiple WSIs, we created a "bag of features" containing the graph representations of all the WSIs associated with the patient. The model is trained using a pairwise ranking loss function to ensure accurate ranking of patients by their methylation statuses.
\begin{equation}
\mathcal{L}=\mathrm{\Sigma}_{p\in B a t c h}\mathrm{\Sigma}_{q\in B a t c h}\max{\left(0,1-\left(F\left(G_p;\theta\right)-F\left(G_q;\theta\right)\right)\right)}  
\end{equation}

\begin{table}[tb]
  \caption{The AUROC and AP scores for the reduced patient cohort for TCGA-GBMLGG for methods SlideGraph$^{\text{methyl}}$,  SlideGraph$^\infty$ \cite{dawood2023cross}, CLAM \cite{lu2021data} and classical machine learning methods \cite{zheng2020whole}.}
  \label{tab:glioma - Reduced dataset}
  \centering
  \begin{tabular}{l|l|l|l}
    \toprule
    Gene group & Method & AUROC & AP\\
    \midrule
    \multirow{8}{*}{Gene Group 0} & SlideGraph$^{\text{methyl}}$ & $\textbf{0.967} \pm \textbf{0.01} $& $\textbf{0.959} \pm \textbf{0.02}$ \\
    & SlideGraph$^\infty$ \cite{dawood2023cross} & $0.954 \pm 0.02 $ & $0.942 \pm 0.02$ \\
    & CLAM \cite{lu2021data} & $0.942 \pm 0.03 $ & $0.925 \pm 0.05$\\
    & Adaboost & $0.520 \pm 0.09$ & $0.529 \pm 0.06$\\
    & LR & $0.532 \pm 0.10$ & $0.537 \pm 0.08$\\
    & Naive Bayes & $0.520 \pm 0.06$& $0.506 \pm 0.05$\\
    & MLP & $0.535 \pm 0.11$& $0.543 \pm 0.08$\\
    & Random Forest &  $0.540 \pm 0.08$& $0.534 \pm 0.07$\\
    & SVM & $0.535 \pm 0.10$& $0.548 \pm 0.05$\\
    \midrule
    
    \multirow{8}{*}{Gene Group 1} & SlideGraph$^{\text{methyl}}$ & $\textbf{0.948} \pm \textbf{0.01}$ & $\textbf{0.949} \pm \textbf{0.01}$ \\
    & SlideGraph$^\infty$ \cite{dawood2023cross} & $0.936 \pm 0.02 $ & $0.933 \pm 0.03$ \\
    & CLAM \cite{lu2021data} & $0.908 \pm 0.03 $& $0.898 \pm 0.03$ \\
    & Adaboost & $0.498 \pm 0.08$ & $0.551 \pm 0.10$\\
    & LR & $0.502 \pm 0.07 $ & $0.540 \pm 0.07$\\
    & Naive Bayes & $0.505 \pm 0.07$& $0.527 \pm 0.08$\\
    & MLP & $0.517 \pm 0.06$& $0.542 \pm 0.08$\\
    & Random Forest & $0.498 \pm 0.07$& $0.519 \pm 0.07$\\
    & SVM & $0.500 \pm 0.08$& $0.514 \pm 0.07$\\
    \midrule

    \multirow{8}{*}{Gene Group 2}& SlideGraph$^{\text{methyl}}$ & $\textbf{0.929} \pm \textbf{0.01}$ & $\textbf{0.922} \pm \textbf{0.02}$ \\
    & SlideGraph$^\infty$ \cite{dawood2023cross} & $0.908 \pm 0.03 $ & $0.891 \pm 0.04$ \\
    & CLAM \cite{lu2021data} & $0.885 \pm 0.02 $ & $0.837 \pm 0.03$ \\
    & Adaboost & $0.901 \pm 0.04$ & $0.873 \pm 0.05$\\
    & LR & $0.907 \pm 0.04 $ & $0.870 \pm 0.04$\\
    & Naive Bayes & $0.809 \pm 0.05$& $0.787 \pm 0.07$\\
    & MLP & $0.869 \pm 0.03$& $0.832 \pm 0.05$\\
    & Random Forest & $0.884 \pm 0.04$& $0.850 \pm 0.06$\\
    & SVM & $0.899 \pm 0.04$& $0.877 \pm 0.04$\\
    \bottomrule
  \end{tabular}
\end{table}

\begin{table}[tb]
  \caption{The AUROC and AP scores of SlideGraph$^{methyl}$, SlideGraph$^\infty$ \cite{dawood2023cross} and CLAM \cite{lu2021data} for the complete patient cohort for TCGA-GBMLGG.}
  \label{tab:glioma-complete}
  \centering
  \begin{tabular}{l|l|l|l}
    \toprule
    Gene group & Method & AUROC & AP\\
    \midrule
    \multirow{2}{*}{Gene Group 0} & SlideGraph$^{\text{methyl}}$ & $\textbf{0.946} \pm \textbf{0.02}$ & $\textbf{0.946} \pm \textbf{0.02}$ \\
    & SlideGraph$^\infty$ \cite{dawood2023cross} & $0.930 \pm 0.03 $ & $0.930 \pm 0.03$ \\
    & CLAM \cite{lu2021data} & $0.923 \pm 0.03 $& $0.921 \pm 0.04$ \\
    \midrule
    \multirow{2}{*}{Gene Group 1} & SlideGraph$^{\text{methyl}}$ & $\textbf{0.941} \pm \textbf{0.02}$ & $\textbf{0.947} \pm \textbf{0.02}$ \\
    & SlideGraph$^\infty$ \cite{dawood2023cross} & $0.923 \pm 0.01 $ & $0.921 \pm 0.02$ \\
    & CLAM \cite{lu2021data} & $0.899 \pm 0.03 $& $0.907 \pm 0.03$\\
    \midrule
    \multirow{2}{*}{Gene Group 2} & SlideGraph$^{\text{methyl}}$ & $\textbf{0.924} \pm \textbf{0.03}$ & $\textbf{0.929} \pm \textbf{0.02}$ \\
    & SlideGraph$^\infty$ \cite{dawood2023cross} & $0.899 \pm 0.03 $ & $0.897 \pm 0.03$ \\
    & CLAM \cite{lu2021data} & $0.892 \pm 0.02 $& $0.891 \pm 0.02$ \\
    \bottomrule
  \end{tabular}
\end{table}

\begin{figure}
\centering
\includegraphics[width=\textwidth]{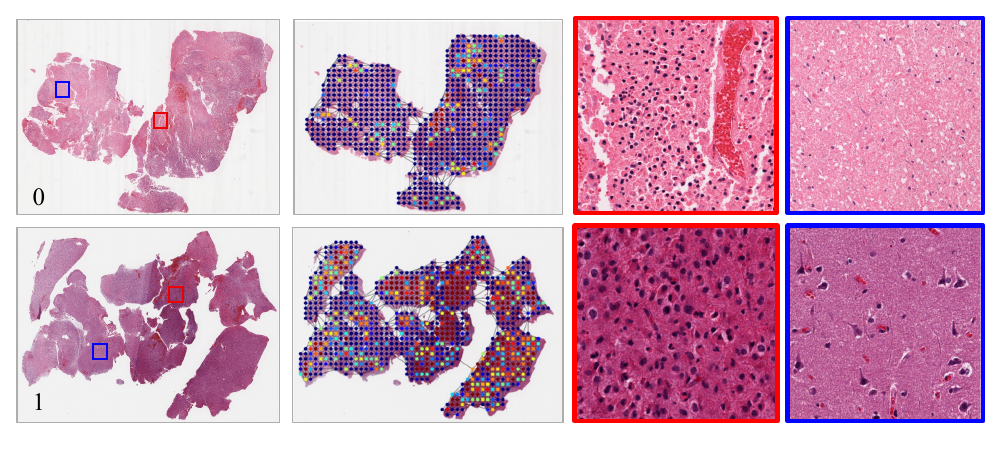}
\caption{Example WSIs for TCGA-GBMLGG gene group 0 for status = 0 (top row) and status = 1 (bottom row) and the corresponding heatmaps. Additionally, we show magnified highly contributing ROIs identified by the proposed method for status = 0 (blue) and status = 1 (red).} \label{glioma_results}
\end{figure}

\subsection{Comparative Analysis}
We compared the proposed workflow with SlideGraph$^\infty$ \cite{dawood2023cross},  Clustering-constrained Attention Multiple instance learning (CLAM) \cite{lu2021data} and with the state-of-the-art (SOTA) methods proposed in the study by Zheng \etal \cite{zheng2020whole} which include Logistic Regression (LR), Support Vector Machine (SVM) and Multi-layer Perceptron classifier (MLP) using 35 cellular morphometric features and eight contextual features. The features were utilized as provided in the official implementation, although it is important to note that these features were not available for the entire patient cohorts.  Consequently, we conducted two distinct sets of experiments, the first on the reduced set of patients for which these features were available, for the sake of a direct comparison with previously published results and the second on the complete set of patients for which feature representations could be extracted. For the TCGA-KIRC dataset, the reduced set comprised 326 patients, whereas the complete set included 511 patients. Similarly, for the TCGA-GBMLGG dataset, the reduced set consisted of 340 patients, while the complete set comprised 729 patients. Furthermore, the SOTA methods by Zheng \etal were applied to the same TCGA cohorts as in our experiments. To ensure fair comparisons, we followed their approach and performed grid search to optimize hyperparameters, as detailed in their code.

\section{Results}
The SlideGraph$^{methyl}$ method was trained for 300 epochs with a mini-batch size of 8 and 3 EdgeConv layers. A learning rate and weight decay of 0.001 and 0.0001 were used with the Adam optimizer. We used stratified five-fold cross-validation at the patient level and the same folds were used for all experiments within each experiment set to ensure a reliable comparison of the performance of all models. 

We use the area under the receiver operating characteristic curve (AUROC) and average precision (AP) as the main metrics of comparison. The gene groups remain the same in both experimental setups.

\subsection{Glioblastoma Multiforme and Brain Lower Grade Glioma}
\label{glioma_results_section}
\begin{figure}[ht]
\centering
\includegraphics[width=\textwidth]{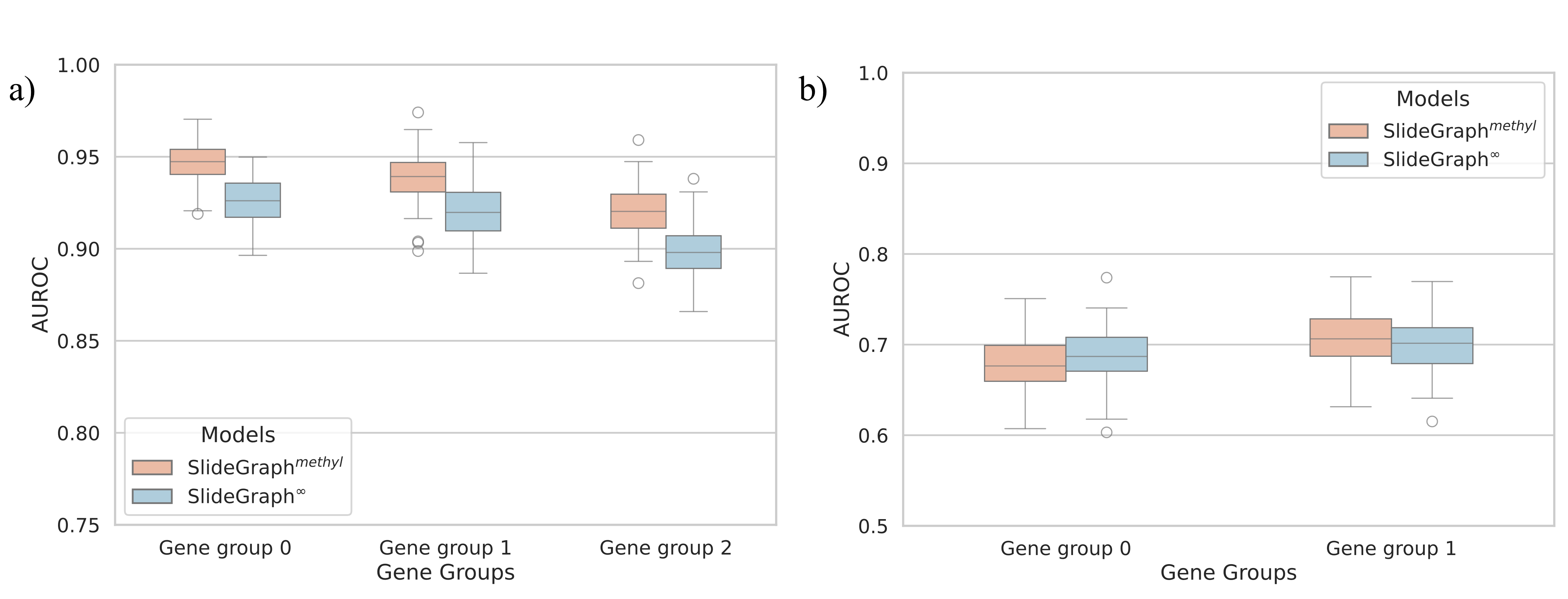}
\caption{ a) Boxplots showing AUROC distribution of SlideGraph$^\infty$ and SlideGraph$^{methyl}$ for the three gene groups across 1,000 bootstrap runs for TCGA-GBMLGG b) Boxplots showing AUROC distribution of SlideGraph$^\infty$ and SlideGraph$^{methyl}$ for the two gene groups across 1,000 bootstrap runs for TCGA-KIRC. } 
\label{bootstrapping}
\end{figure}

As shown in Table.\ref{tab:glioma - Reduced dataset} in the reduced patient cohort, the proposed method demonstrated significantly better performance compared to the SOTA machine learning methods, particularly within gene groups 0 and 1. Notably, the SlideGraph$^{methyl}$ workflow outperformed all other methods across both sets of experiments, achieving the highest AUROC scores of (0.97, 0.95 and 0.93) and (0.95, 0.94 and 0.92) and AP scores of  (0.96, 0.95 and 0.92) and (0.95, 0.95 and 0.93), respectively, as illustrated in Tables.\ref{tab:glioma - Reduced dataset} and \ref{tab:glioma-complete}. In Fig.\ref{bootstrapping} a. we illustrate the AUROC distributions of the two best performing methods, SlideGraph$^{methyl}$ and SlideGraph$^\infty$ methods across 1,000 bootstrap runs on the complete patient cohort (\textit{p} $<0.01$). We note that the proposed method has a higher median AUROC across all three gene groups. Moreover, we generated spatially enriched graph-based heatmaps using the proposed method by visualizing the node-level prediction scores from WSIs for gene group 0 (status 0 and 1) as depicted in Fig.\ref{glioma_results}. Regions highlighted in red indicate an association with status = 1, while regions highlighted in blue translate to an association with status = 0.  We note that the regions outlined in red exhibit relatively high cellular density and more intense staining as compared to the regions outlined in blue.

In order to analyze the clinical and pathological significance of the gene groups, we performed gene set enrichment analysis (GSEA) on the genes belonging to each group using Enrichr \cite{kuleshov2016enrichr}. We obtained a list of enriched terms and their adjusted \textit{p}-values from the Molecular Signatures Database (MSigDB) hallmark 2020 library as illustrated in Figs.\ref{gse} a,b,c. To establish statistical significance, we applied a cutoff value of $\textit{p} < 0.05$ on the adjusted \textit{p}-values. Notably, group 0 showed enrichment for phosphatidylinositol 3-kinase (PI3K)/AKT/ mechanistic target of rapamycin (mTOR) signaling which is frequently dysregulated in cancers and has been associated with disease progression in IDH-mutant diffuse gliomas \cite{mohamed2022pi3k}. Both groups 0 and 2 showed enrichment for Epithelial Mesenchymal Transition (EMT) which significantly contributes to the high invasive-ness of gliomas. During EMT, glioma cells transition to a mesenchymal phenotype, enhancing their invasive and migratory capabilities \cite{xing2022emerging}. Both groups 1 and 2 showed enrichment for Hypoxia (insufficient oxygenation) which has been associated with increased invasion and aggression in Glioblastoma (GBM), resulting in poor patient outcomes. \cite{marallano2024hypoxia}.  

\begin{table}[tb]
  \caption{The AUROC and AP scores for the reduced patient cohort for TCGA-KIRC for methods SlideGraph$^{\text{methyl}}$,  SlideGraph$^\infty$ \cite{dawood2023cross}, CLAM \cite{lu2021data} and classical machine learning methods \cite{zheng2020whole}.}
  \label{tab:rcc-reduced dataset}
  \centering
  \begin{tabular}{l|l|l|l}
    \toprule
    Gene group & Method & AUROC & AP\\
    \midrule
    \multirow{8}{*}{Gene Group 0} & SlideGraph$^{\text{methyl}}$ & $\textbf{0.686} \pm \textbf{0.01}$ & $\textbf{0.736} \pm \textbf{0.04}$ \\
    & SlideGraph$^\infty$ \cite{dawood2023cross} & $0.659 \pm 0.02 $ & $0.720 \pm 0.02$ \\
    & CLAM \cite{lu2021data} & $ 0.639 \pm 0.07 $& $0.709 \pm 0.07$\\
    & Adaboost & $0.487 \pm 0.07$ & $0.590 \pm 0.10$\\
    & LR & $0.488 \pm 0.05$ & $0.608 \pm 0.09$ \\
    & Naive Bayes & $0.471 \pm 0.05$ & $0.583 \pm 0.08$ \\
    & MLP & $0.455 \pm 0.08 $ & $0.567 \pm 0.10$ \\
    & Random Forest & $0.462 \pm 0.07$ & $0.592 \pm 0.12$ \\
    & SVM & $0.469 \pm 0.05$ & $0.593 \pm 0.11$\\
    \midrule
    
    \multirow{8}{*}{Gene Group 1} & SlideGraph$^{\text{methyl}}$ & $0.720 \pm 0.05$ & $\textbf{0.608} \pm \textbf{0.07}$ \\
    & SlideGraph$^\infty$ \cite{dawood2023cross} & $0.699 \pm 0.07 $ & $0.580 \pm 0.08$ \\
    & CLAM \cite{lu2021data} & $ 0.687 \pm 0.04 $& $0.575 \pm 0.06$\\
    & Adaboost & $0.628 \pm 0.07$ & $ 0.493 \pm 0.07$ \\
    & LR & $0.630 \pm 0.08 $ & $0.516  \pm 0.12$ \\
    & Naive Bayes & $0.639 \pm 0.05$ & $0.507 \pm 0.09 $ \\
    & MLP & $\textbf{0.735} \pm \textbf{0.03} $ & ${0.585} \pm {0.06}$ \\
    & Random Forest & $0.692 \pm 0.03$ & $0.559 \pm 0.08$ \\
    & SVM & $ 0.694 \pm 0.05$ & $0.564 \pm 0.11$ \\
    \bottomrule
  \end{tabular}
\end{table}

\begin{table}[tb]
  \caption{The AUROC and AP scores of SlideGraph$^{\text{methyl}}$,  SlideGraph$^\infty$ \cite{dawood2023cross} and CLAM \cite{lu2021data} for complete patient cohort for TCGA-KIRC. }
  \label{tab:rcc-complete}
  \centering
  \begin{tabular}{l|l|l|l}
    \toprule
    Gene group & Method & AUROC & AP \\
    \midrule
    \multirow{2}{*}{Gene Group 0} & SlideGraph$^{\text{methyl}}$ & $0.676 \pm 0.02$ & $0.730 \pm 0.02$ \\
    & SlideGraph$^\infty$ \cite{dawood2023cross} & $\textbf{0.686} \pm \textbf{0.03} $ & $\textbf{0.750} \pm \textbf{0.03}$ \\
    & CLAM \cite{lu2021data} & $0.647 \pm 0.08$& $0.721 \pm 0.09$\\
    
    \midrule
    \multirow{2}{*}{Gene Group 1} & SlideGraph$^{\text{methyl}}$ & $\textbf{0.710} \pm \textbf{0.05}$ & $\textbf{0.611} \pm \textbf{0.07}$ \\
    & SlideGraph$^\infty$ \cite{dawood2023cross} & $0.700 \pm 0.01 $ & $0.585 \pm 0.05$ \\
    & CLAM \cite{lu2021data} & $0.699 \pm 0.05$& $0.596 \pm 0.07$\\
    \bottomrule
  \end{tabular}
\end{table}

\begin{figure}[ht]
\centering
\includegraphics[width=\textwidth]{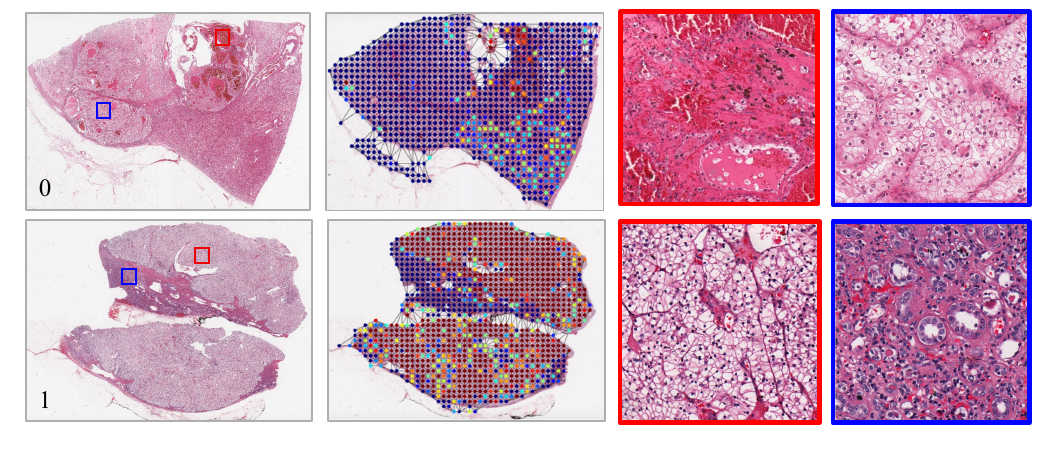}
\caption{Example WSIs for TCGA-KIRC gene group 0 for status = 0, (top row) and status = 1, (bottom row) and the corresponding heatmaps. Additionally, we show magnified highly contributing ROIs identified by the proposed method for status = 0 (blue) and status = 1 (red).} 
\label{rcc_results}
\end{figure}

\subsection{Kidney Renal Clear Cell Carcinoma}
In the reduced patient cohort for TCGA-KIRC, the proposed method outperformed the SOTA classical machine learning approaches in group 0 (0.69 and 0.74) and achieved competitive results for group 1 as shown in Table.\ref{tab:rcc-reduced dataset}. The superior performance of the MLP for AUROC in group 1 may be due to clearer class separation in this group. However, it's lower performance for the AP metric indicates a sensitivity to the imbalanced distribution in group 1 (Label 0 $n=658$, Label 1 $n=410$). Conversely, in group 0, where the data may present more complex relationships, the MLP's performance declined, likely due to the need for deeper spatial feature learning, which GNNs are better equipped to handle. The highest AUROC and AP scores were achieved by the SlideGraph$^{methyl}$ method for group 1 in the complete patient cohort (0.71 and 0.61) as illustrated in Table. \ref{tab:rcc-complete} and competitive performance for group 0. In Fig.\ref{bootstrapping} b. we present box-plots that display the AUROC distributions of the SlideGraph$^{methyl}$ and SlideGraph$^\infty$ across 1,000 bootstrap runs on the complete patient cohort for gene group 0 and gene group 1 (\textit{p} $<0.01$). Fig.\ref{rcc_results}. illustrates graph-based heatmaps generated using the proposed method from WSIs of the TCGA-KIRC dataset for group 0 (status 0 and 1).  

\begin{figure}[ht]
\centering
\includegraphics[width=\textwidth]{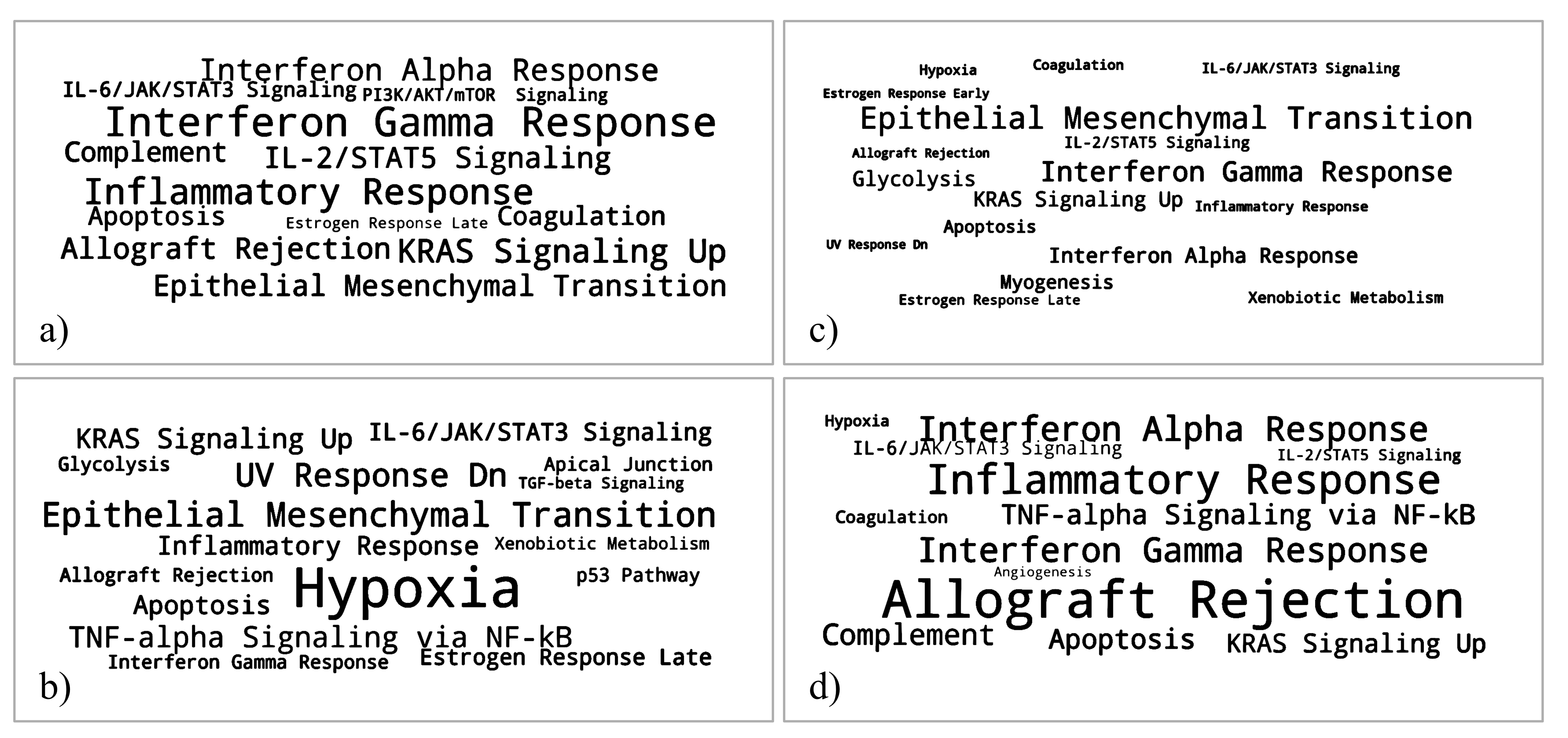}
\caption{Enriched terms for hallmark processes in a) ,b) and c) for gene groups 0, 1 and 2, respectively, for TCGA-GBMLGG and d) gene group 0 for TCGA-KIRC. The font sizes are proportional to the significance of the enriched term based on the adjusted \textit{p}-values.} 
\label{gse}
\end{figure}

We also performed GSEA on genes belonging to each group for the TCGA-KIRC dataset and show the statistically significant enriched terms for group 0 in Fig.\ref{gse} d. These included inflammatory response, angiogenesis and hypoxia. Inflammation plays a key role in advanced renal cell carcinoma, facilitating tumor progression and metastasis by interacting with the tumor microenvironment and immune cells \cite{kruk2023inflammatory}. Angiogenesis is a critical process for cancer growth and metastasis plays a vital role in the prognosis of kidney renal clear cell carcinoma  \cite{zheng2021multi}. Hypoxia has been shown to contribute to tumor angiogenesis and has been associated with poor prognosis \cite{chappell2019hypoxia}. When analyzing the genes in group 1, it was found that they did not show statistically significant enrichment in the MSigDB Hallmarks 2020 dataset. These genes might be involved in more subtle or less well-defined biological roles or require further investigation. 

\section{Discussion and Conclusion}
In this study, we proposed a deep learning based computational workflow to predict gene group level DNA methylation states for each patient using histopathology images. To the best of our knowledge, ours is the first method to use graph-based weakly-supervised learning for the prediction of group-level methylation states across multiple cancer types. We achieved significantly better results than previous SOTA methods. This underscores the importance of employing feature encoders pre-trained on histopathology images. 

We note that our results on the TCGA-GBMLGG dataset are higher than those on the TCGA-KIRC dataset. This may be due to the brain having among the highest levels of DNA methylation of any body tissue, thereby having a greater impact on tissue morphology \cite{moore2013dna}. This is further supported by the fact that fewer genes were identified as differentially methylated in TCGA-KIRC as compared to TCGA-GBMLGG. We performed gene set enrichment analysis on the gene groups and discovered statistically significant associations with known cancer hallmark processes for majority of the gene groups. Moreover, we generated spatially enriched graph-based heatmaps to analyze the connections between visual patterns in histology images and DNA methylation patterns. 

Our study shows that DNA methylation states of cancer-related genes can be accurately predicted from H\&E WSIs using GNN models. Traditional methods of assessing DNA methylation are typically complex and time-consuming. Our approach leverages the rich visual information contained in WSIs, which are routinely used in clinical settings, to predict these states. Our method provides insights into the epigenetic landscape of tumors and can help us understand the role of DNA methylation changes in identifying digital biomarkers, which could potentially be used for early diagnosis and detection of diseases. In future, we plan to enhance this method using multiple input modalities and to extend the proposed workflow to multiple cancer types.

\section*{Acknowledgements}
The authors acknowledge support from AstraZeneca, UK and the Department of Computer Science at the University of Warwick, UK.

%
%
\bibliographystyle{splncs04}
\bibliography{main}
\end{document}